\documentclass[letter]{aa}  

\usepackage{graphicx}
\usepackage{txfonts}
\usepackage{amssymb}
\usepackage{lineno}
\usepackage{txfonts}
\usepackage{natbib}   
\usepackage{amstext}
\usepackage{mathabx}
\usepackage{multirow}
\usepackage{ulem}
\usepackage{color}
\usepackage{pifont}
\usepackage{lscape}
\usepackage{mathtools}
\usepackage{enumitem}

\usepackage{hyperref}
\newcommand{\science}{Science}
\newcommand{\camcos}{Commun. Appl. Math. Comput. Sci.} 
\newcommand{\joss}{J. Open Source Softw.} 
\newcommand{\njp}{New J. Phys.} 
\newcommand{\jatmos}{J. Atmos. Science} 
\newcommand{\natastron}{Nat. Astron.} 

\newcommand{\fig}[1]{Fig.~\ref{#1}}

\newcommand{\degre}{\ensuremath{^\circ}}

\begin{document} 


   \title{First direct measurement of auroral and equatorial jets in the stratosphere of Jupiter}


   \author{T.~Cavali\'e\inst{1,2}
          \and
          B.~Benmahi\inst{1}
          \and
          V. Hue\inst{3}
          \and
          R.~Moreno\inst{2}
          \and
          E.~Lellouch\inst{2}
          \and
          T.~Fouchet\inst{2}
          \and
          P.~Hartogh\inst{4}
          \and
          L.~Rezac\inst{4}
          \and
          T.~K.~Greathouse\inst{3}
          \and
          G.~R.~Gladstone\inst{3}
          \and
          J.~A.~Sinclair\inst{5}
          \and
          M.~Dobrijevic\inst{1}
          \and
          F.~Billebaud\inst{1}
          \and
          C.~Jarchow\inst{4}
          }

   \institute{Laboratoire d'Astrophysique de Bordeaux, Univ. Bordeaux, CNRS, B18N, all\'ee Geoffroy Saint-Hilaire, 33615 Pessac, France\\
              \email{thibault.cavalie@u-bordeaux.fr}
         \and
             LESIA, Observatoire de Paris, PSL Research University, CNRS, Sorbonne Universit\'es, UPMC Univ. Paris 06, Univ. Paris Diderot, Sorbonne Paris Cit\'e, Meudon, France\\
          \and
             Southwest Research Institute, San Antonio, TX 78228, United States\\
          \and
             Max-Planck-Institut f\"ur Sonnensystemforschung, 37077 G\"ottingen, Germany\\
          \and
             Jet Propulsion Laboratory, California Institute of Technology, 4800 Oak Grove Drive, Pasadena, CA 91109, USA\\
             }

   \date{Received January 12, 2021; Accepted February 6, 2021}

 
  \abstract
   {The tropospheric wind pattern in Jupiter consists of alternating prograde and retrograde zonal jets with typical velocities of up to 100\,m/s around the equator. At much higher altitudes, in the ionosphere, strong auroral jets have been discovered with velocities of 1-2\,km/s. There is no such direct measurement in the stratosphere of the planet.}
   {In this paper, we bridge the altitude gap between these measurements by directly measuring the wind speeds in Jupiter's stratosphere. }
   {We use the Atacama Large Millimeter/submillimeter Array's very high spectral and angular resolution imaging of the stratosphere of Jupiter to retrieve the wind speeds as a function of latitude by fitting the Doppler shifts induced by the winds on the spectral lines.}
   {We detect for the first time equatorial zonal jets that reside at 1\,mbar, i.e. above the altitudes where Jupiter's Quasi-Quadrennial Oscillation occurs. Most noticeably, we find 300-400\,m/s non-zonal winds at 0.1\,mbar over the polar regions underneath the main auroral ovals. They are in counter-rotation and lie several hundreds of kilometers below the ionospheric auroral winds. We suspect them to be the lower tail of the ionospheric auroral winds.}
   {We detect directly and for the first time strong winds in Jupiter's stratosphere. They are zonal at low-to-mid latitudes and non-zonal at polar latitudes. The wind system found at polar latitudes may help increase the efficiency of chemical complexification by confining the photochemical products in a region of large energetic electron precipitation.}


   \maketitle

\section{Introduction}
The tropospheric zonal wind system of Jupiter has been observed for decades showing alternating prograde and retrograde jets at the boundaries between zones and belts \citep{Chapman1969,Ingersoll1979,Ingersoll2004,Limaye1982,Garcia-Melendo2001}. A similar structure, although with jets fewer in number, has also been found in Saturn's troposphere \citep{Smith1981,Sanchez-Lavega2000,Choi2009,Barbara2021}. The mechanism behind the winds as well as their vertical extent has been extensively studied. Recent Juno and Cassini gravity field measurements have demonstrated that these winds extend down to a few thousands of kilometers below the cloud deck in Jupiter and Saturn and are therefore powered by the internal heat flux \citep{Kaspi2018,Kaspi2020,Guillot2018,Galanti2019}. 

Above the tropopause, in the stratosphere, there are no tracers to infer the wind pattern from visible light imaging. The stratospheric winds could only be derived in the 20-50\,mbar and 30\degre S-60\degre S ranges on the exceptional occasion of the Shoemaker-Levy 9 impacts from the evolution of the debris fields \citep{Banfield1996,Sanchez-Lavega1998}. The relative contributions of thermal versus mechanical forcing (by e.g. waves and eddies) in the stratosphere are therefore unquantified. So far, the stratospheric zonal wind pattern has only been indirectly derived from the thermal wind balance relation applied to the measured zonal temperature field. There are several studies that applied this method to Jupiter and Saturn \citep{Flasar2004,Flasar2005,Fouchet2008,Guerlet2011,Guerlet2018,Fletcher2016,Cosentino2017}. In addition, systematic infrared observations of long-time scales led to the discovery of stratospheric quasi-periodic oscillations manifested in the stratospheric temperatures and winds, in particular the Quasi-Quadrennial Oscillation (QQO) in Jupiter \citep{Orton1991} and also the Saturn Equatorial Oscillation (SEO - \citealt{Orton2008}). These oscillations have been the subject of numerous follow-up observations and modeling efforts to obtain robust constraints on their origin and evolution \citep{Cosentino2017,Li2000,Medvedev2013,Spiga2020,Bardet2021,Giles2020,Antunano2020}. However, deriving the wind field from the thermal wind balance is only an approximation which unfortunately breaks down at the equator, even if a new prescription of this equation for equatorial latitudes has been recently proposed \citep{Marcus2019}. Solving the thermal wind equation also requires a boundary condition, often taken as the cloud-top wind pattern. Furthermore, the temperature field is only interpolated between the tropopause and the middle stratosphere where it can be retrieved from hydrocarbon emissions. In any case, the thermal wind equation only gives wind shear but not the absolute wind speeds. The real magnitude of the stratospheric winds has thus remained elusive. Direct wind measurements in the stratosphere are thus warranted to quantify the role of thermal and mechanical forcing, and thus better constrain models of the planetary wave propagation that generate the stratospheric equatorial oscillations.

With spectral resolving powers, $R=\lambda/\Delta\lambda$, exceeding 10$^6$, heterodyne spectroscopy in the millimeter wavelength range has opened the possibility to directly measure frequency Doppler shifts induced by winds in spectral lines of molecular species, as originally demonstrated at Venus and Mars \citep{Shah1991,Lellouch1991}. The Atacama Large Millimeter/submillimeter Array (ALMA) now enables almost instantaneous mapping, with high sensitivity, and sufficiently high spectral and angular resolutions to measure wind-induced Doppler shifts in most Solar System atmospheres (e.g. \citealt{Lellouch2019}). At Jupiter, the main difficulty resides in measuring Doppler shifts caused by $\sim$100\,m/s (or less) winds superimposed onto the rapid Jovian rotation (12.5\,km/s at the equator). To overcome this challenge, we use the strong millimeter lines of HCN and CO, two species delivered by the impacts of comet Shoemaker-Levy 9 (SL9) in 1994 \citep{Lellouch1995}. The SL9-derived species were expected to be homogeneously distributed in latitude at the time of our observations (e.g. \citealt{Moreno2003,Cavalie2013}).

\section{Observations \label{Observations}}
We observed Jupiter with the ALMA interferometer on March 22$^\mathrm{nd}$, 2017, at 5:11-5:36UT with 42 12-m antennas, as part of the 2016.1.01235.S project. At this time, Jupiter's equatorial diameter subtended a 43.8\arcsec~angle, the sub-earth point latitude was -3\degre, and the central meridian longitude (CML) ranged from 65\degre~to 80\degre~(System III). To map the whole planet at such frequencies, we had to use a mosaic of 39 pointings. Standard pointing, bandpass, amplitude and phase calibration observations were carried out and accounted for in the data reduction we performed under CASA 4.7.2 (additional details can be found in Section~\ref{appendixA}). The lack of short-spacings with the interferometer results in filtering out Jupiter's extended emission (i.e. most of the disk flux), such that only the limb observations are preserved. The baselines of the interferometer ranged from 15.1 to 160.7\,m, providing an elliptical synthesized beam of 1.2\arcsec (East-West) $\times$ 1\arcsec (North-South). This resulted in a latitudinal resolution of $\sim$3\degre~at the equator, degrading to $\sim$10\degre~close to the poles. From each spectral cube, we extracted $\sim$550 spectra located at the planet limb (at the 1\,bar level) to oversample the beam by a factor of 4 to 5. The accumulated on-source integration time of 24 minutes enabled us to detect the HCN (5-4) and CO (3-2) emissions at 354.505\,GHz and 345.796\,GHz (respectively) with signal-to-noise ratios (S/N) of $\sim$25 per beam at the limb at spectral resolutions of 122\,kHz and 488\,kHz (respectively).

\section{HCN and CO vertical and horizontal distributions}
\begin{figure*}[!h]
\begin{center}
   \includegraphics[width=14cm,keepaspectratio]{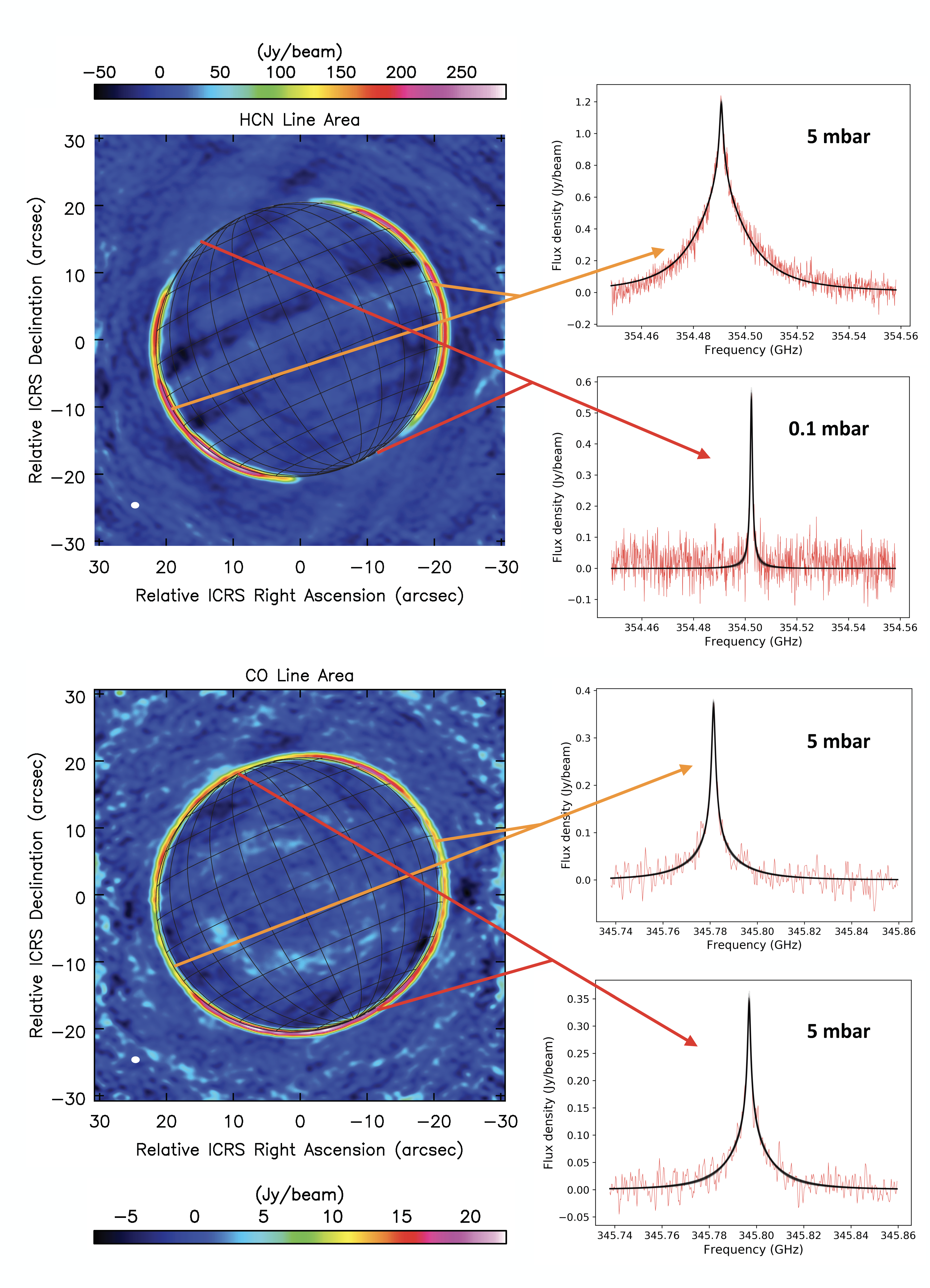}
\end{center}
\caption{ALMA observations of Jupiter's stratospheric HCN and CO. (Left) Line area maps of the HCN (5-4) (top) and CO (3-2) (bottom) emission at the limb of Jupiter. (Right) Spectra extracted from the data cubes (red lines) showing typical line shapes and the cut-off pressure ($p_0$) in the species vertical profile to reproduce the line width, with the 30 best-fit spectra computed with the MCMC procedure from the parametrized line shape. Observable Doppler shifts with respect to the lines rest frequencies are caused by the planet's rapid rotation and the local east-west winds. }
\label{Map_spectra} 
\end{figure*}

The spectral lines show limited variability in terms of amplitude, but the HCN lines present some variability in terms of width. We analyzed the vertical distributions of CO and HCN as a function of latitude from the line widths. Using empirical vertical profiles of CO and HCN with a cut-off pressure $p_0$ below which the species have a constant mole fraction, and the radiative transfer model of \citet{Cavalie2019}, we found that CO is present at $p_0$ $<$ 5\,mbar at all latitudes, whereas HCN is found at the same pressure levels only at the low-to-mid latitudes (60\degre S-50\degre N). At higher latitudes, HCN is restricted to $p_0$ $<$ 0.1\,mbar (see \fig{Map_spectra}). This is surprising because HCN and CO share the same origin, are both long-lived, and should thus have similar horizontal and vertical distributions. The missing spectral signature of HCN at pressures higher than 0.1\,mbar and at high latitudes exhibits asymmetry in latitude between the northern and the southern hemispheres: the transition between the broad HCN lines seen at low and mid-latitudes with the thin HCN lines seen in the polar region is at 60\degre S versus 50\degre N. These facts point to a chemical sink for HCN related to the aurorae, which latitudinal extent shows similar asymmetry in latitude between the north and the south. In particular, aerosols are known to be more abundant at high latitudes \citep{Zhang2013}, suggesting adsorption of HCN on aurorally-produced aerosols as a potential sink mechanism \citep{Anderson2016}.

\section{Wind speed retrieval}
\begin{figure*}[!h]
\begin{center}
   \includegraphics[width=15cm,keepaspectratio]{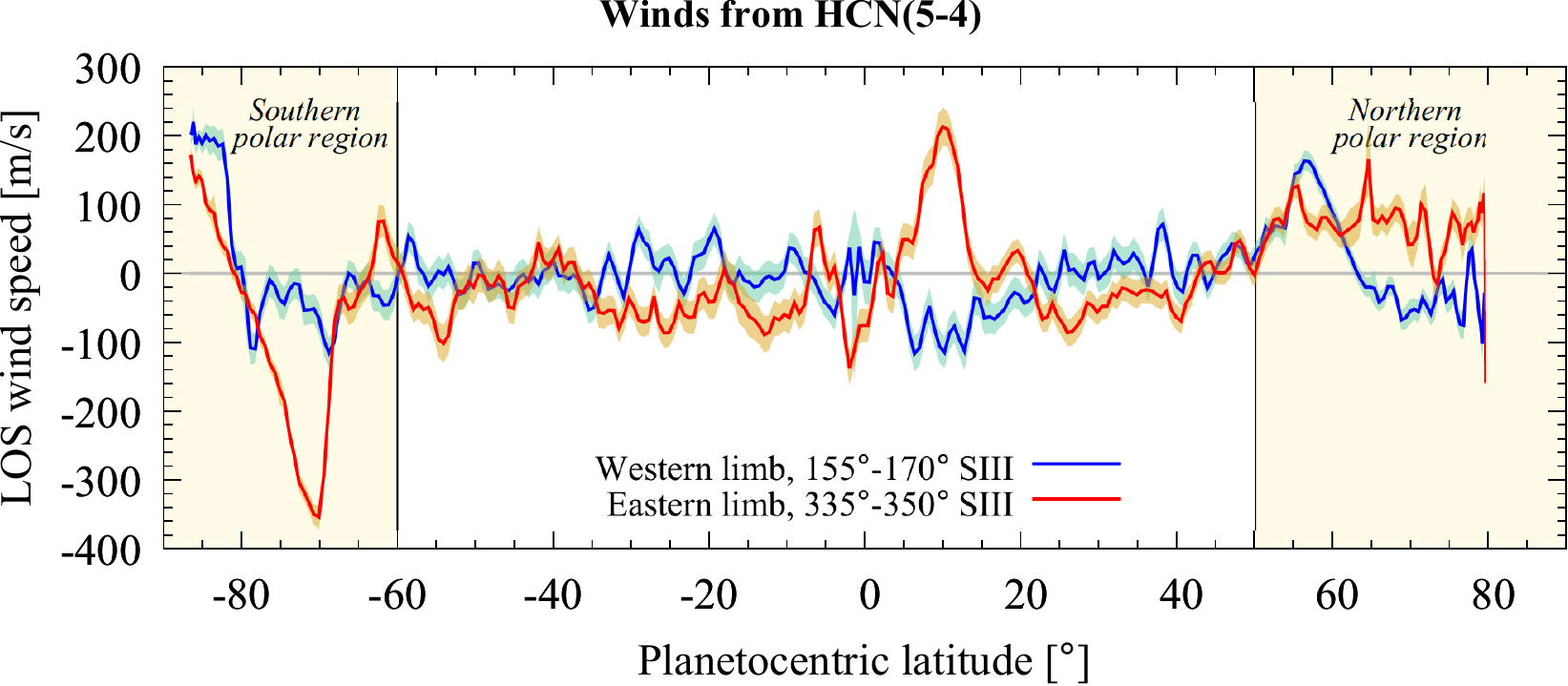}
    
   \vspace{0.5cm}
      
   \includegraphics[width=15cm,keepaspectratio]{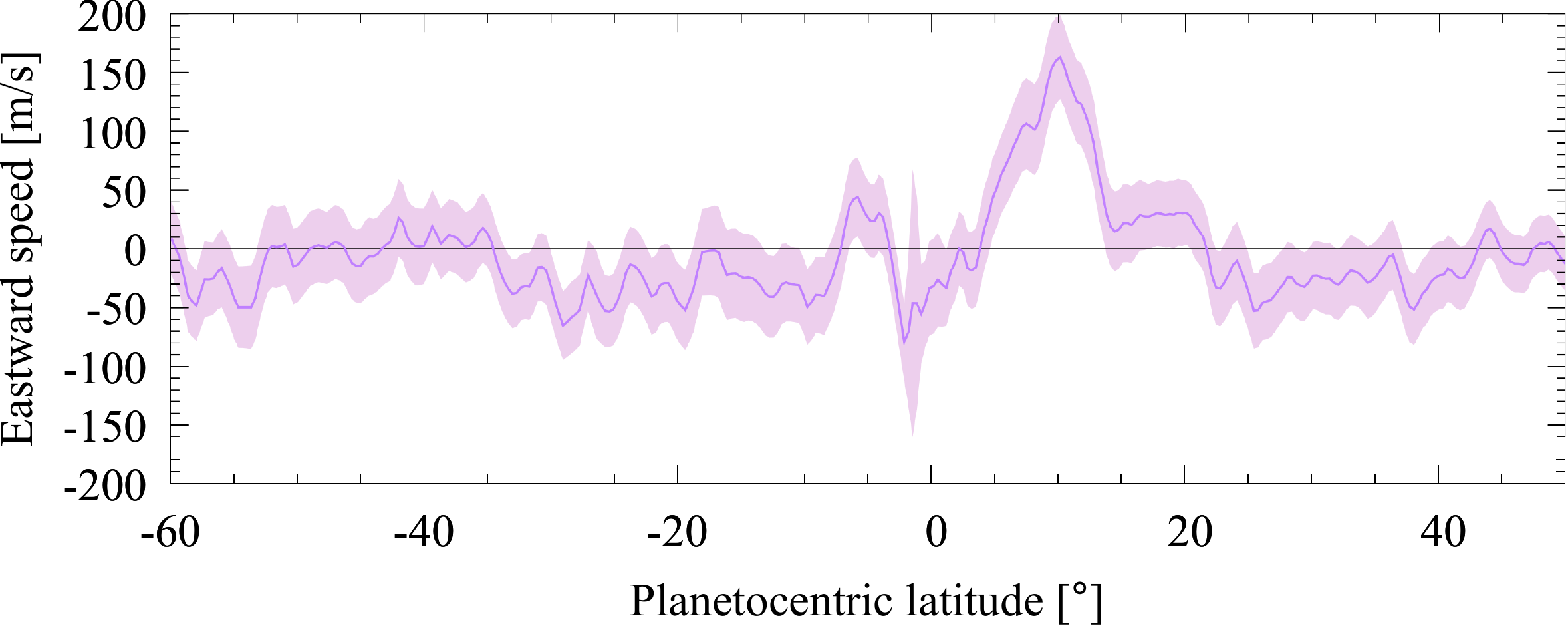}
\end{center}
\caption{Jupiter's stratospheric winds. (Top) Instantaneous line-of-sight wind speed measurements as a function of latitude obtained from ALMA spectral mapping observations of the HCN (5-4) line. The winds are measured at 1\,mbar from 60\degre S to 50\degre N and at 0.1\,mbar at polar latitudes. The western limb data is plotted in blue and the eastern limb data in red (1-$\sigma$ uncertainty envelopes in light blue and orange, respectively). (Bottom) Eastward wind speeds averaged from both limbs from 60\degre S to 50\degre N. Prograde winds have positive speed values. }
\label{winds_summary} 
\end{figure*}

Within a synthetic beam, the line is naturally Doppler-shifted by the rapid rotation of the planet. Any additional Doppler shift of the line is then indicative of atmospheric motions along the line-of-sight located at the altitude of the wind. ``Wind contribution functions'', as defined by \citet{Lellouch2019}, indicate that fitting the HCN line enables us to retrieve wind speeds at $\sim$1\,mbar from 60\degre S to 50\degre N and at 0.1\,mbar at polar latitudes (see Section~\ref{appendixB} and \fig{S1}).  We determined the line-of-sight (LOS) wind speeds as a function of latitude by fitting the HCN lines with a Markov Chain Monte Carlo (MCMC) scheme \citep{Goodman2010,Foreman-Mackey2019}. We fitted all extracted limb spectra using a parametrized line shape that is fully defined by four parameters (see Section~\ref{appendixC} for a more detailed description). The only parameter of interest is the central frequency of the line; the other three parameters are related to the width and amplitude of the line and help us having a good rms metric for the fitting algorithm. This method is independent of any prior knowledge of the HCN and CO distributions, and atmospheric temperature. We use several hundreds of iterations to fit the line center position and derive its uncertainty. The altitudes of the winds are estimated from the contribution function calculations, as described above. \fig{winds_summary} (top) shows the derived LOS wind speeds as a function of latitude we obtain from HCN. The associated uncertainties result from the combination of the continuum subtraction on the spectra, uncertainties in the subtraction of the planet rotation associated with pointing errors, and uncertainty of the MCMC fitting procedure (see Section~\ref{appendixD}). The wind speeds in \fig{winds_summary} are measured instantaneously, as opposed to the zonal mean wind speeds at the cloud-top found in the literature (e.g. \citealt{Ingersoll2004}). The combination of lower spectral resolution and S/N of the CO observations does not allow us to retrieve wind speeds. We can only put a 3-sigma upper limit of 150\,m/s at 3\,mbar, where fitting the CO line would enable us to measure wind speeds.

\section{Results}
\subsection{Wind speed retrieval results at low-to-mid latitudes}
From 60\degre S to 50\degre N, the strongest and broadest wind we detect is located at 9-11\degre N, as shown in \fig{winds_summary}. It is a prograde jet with a peak LOS velocity of +215$\pm$25\,m/s on the planet eastern limb and -115$\pm$25\,m/s on the planet western limb. The average eastward wind speed is 165$\pm$40\,m/s (\fig{winds_summary} bottom), compatible with the magnitude of the near-equatorial jet found from the thermal wind balance by \citet{Flasar2004} This jet has a full-width at half maximum (FWHM) of $\sim$7\degre. The difference in peak velocities between the two limbs indicates that the local vortices could accelerate/decelerate winds by $\sim$50\,m/s. This situation could thus be similar to what is seen at the cloud level, where \citet{Garcia-Melendo2011} found that the equatorial zone shows variability in wind speeds of $\sim$20\,m/s on average (but up to 60\,m/s) over only one planet rotation, because of vortices and planetary waves. We tentatively find a retrograde jet at 2\degre S with a 2-$\sigma$ confidence level only. Its speed on the eastern limb is -140$\pm$25\,m/s, but we cannot unambiguously identify it on the western limb. The presence of a prograde jet at 4-7\degre S is even more tentative (1.5-$\sigma$). The equatorial wind structure at 1\,mbar is thus asymmetrical with respect to the equator, contrary to the cloud-top wind structure and contrary to what one would expect in the QQO altitude and latitude ranges. It may result from the latitudinal temperature gradients found between the upper stratospheric layers and the mbar region where the QQO occurs \citep{Cosentino2017}, which are also found asymmetrical at the time of our observations \citep{Giles2020}. In the northern and southern low-to-mid latitude, there is little evidence of other jets outside the equatorial region.

\subsection{Wind speed retrieval results in the polar regions}
\begin{figure*}[!h]
\begin{center}
   \includegraphics[width=15cm,keepaspectratio]{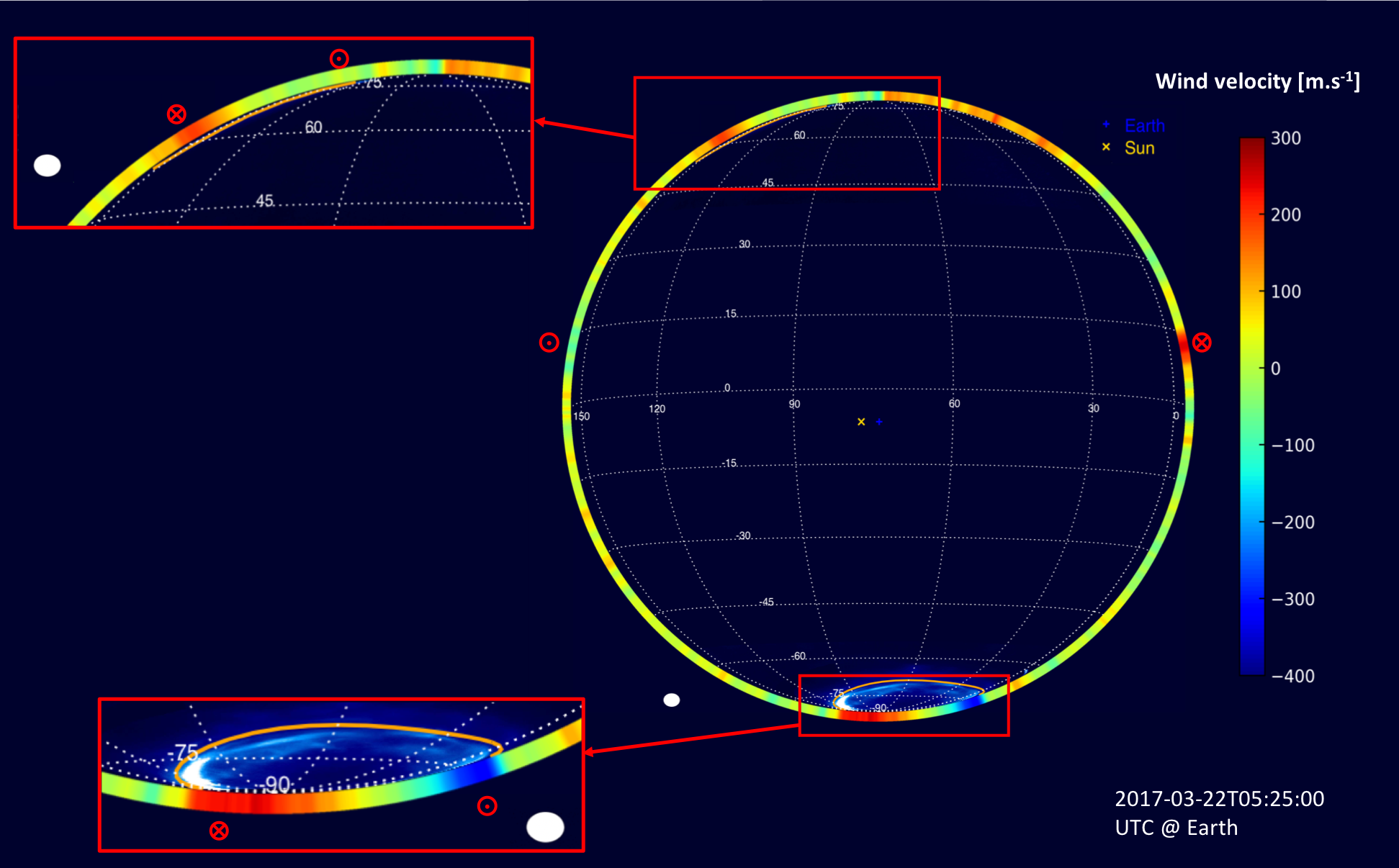}
\end{center}
\caption{Jupiter's UV aurora and stratospheric HCN winds. Composite image showing the LOS wind velocities in\,m/s derived from the ALMA observations and the statistical emission of the aurorae \citep{Clarke2009} in the configuration of the ALMA observations. The northern and southern aurora regions are best seen in dedicated zoomed-in quadrants. The M=30 footprints of the magnetic field model of \citet{Connerney2018} is a good marker of the position of the main ovals as seen by Juno-UVS \citep{Gladstone2017} and is plotted in orange. The white ellipses indicate the spatial resolution of the ALMA observations. The directions of the strongest winds in the equatorial and auroral regions are indicated with the $\odot$ and $\otimes$ red symbols. }
\label{ALMA_comparison} 
\end{figure*}

The most unexpected and outstanding feature detected in our observations are the non-zonal winds seen in the northern and southern polar regions (see \fig{winds_summary} top). We detect jets at 0.1\,mbar at 55\degre N and 85\degre S on the western limb, and at 70\degre S on the eastern one. They all have counter-rotation velocities. The strongest one, seen at 70\degre S on the eastern limb, has an FWHM of 7\degre~and a peak LOS velocity of -350$\pm$20\,m/s. The wind seen at 85\degre S on the western limb peaks at +200$\pm$20\,m/s. These peaks seem to be collocated with the position of the southern auroral oval for the CML of our observations, when compared with the position of the statistical emission of the aurorae \citep{Clarke2009} and with the M=30 footprints of \citet{Connerney2018}'s model of the magnetic field (i.e., the footprints of field lines that reach 30 Jupiter radii at the equator). The latter is a good marker of the position of the main ovals as observed by Juno's Ultraviolet Spectrograph (UVS - \citealt{Gladstone2017}). This comparison can be seen qualitatively in \fig{ALMA_comparison}. The wind peaks at 70\degre S on the eastern limb and at 85\degre S on the western limb would then result from the same jet. To confirm this finding, we implemented a model in which we assumed a constant wind within the southern oval and no wind outside the oval. We took the inner and outer oval edges as defined by \citet{Bonfond2012}. We simulated spectra at infinite spatial and spectral resolutions, Doppler-shifted them according to the LOS auroral oval wind component after carefully accounting for the geometry of the observations, and finally convolved them to the spectral and spatial resolutions of the ALMA observations. To improve the fit, we had to extend by $\sim$2\degre~the inner and outer edges of the southern oval. This model demonstrates that a 370\,m/s counter-rotation wind inside the auroral oval results in asymmetric components as observed at 70\degre S and 85\degre S (see Section~\ref{appendixE} and \fig{S3}). However, this simple model is unable to fit properly the wind speeds within the entire auroral region. The real wind pattern in the auroral region is certainly more complex than in our simple model, like the ionospheric wind field derived by \citet{Johnson2017} from H$_3^+$ emission in the northern auroral region. The lack of spatial resolution prevents us from refining it further without additional and unconstrained parametrization (e.g. variable wind speed within the oval, wind gradient at the interface between the oval and its surroundings, winds not only limited to the oval but also inside the auroral regions). 

It is noteworthy we find hints of a similar counter-rotation jet in the northern auroral region with peak LOS velocities of +165$\pm$15\,m/s and a FWHM of 6\degre~in latitude at 57\degre N on the western limb. The northern oval was just coming into view at the time of the observations, thus severely limiting the viewing of the northern auroral region. A significant part of the main oval was expected to be close to tangential to the limb on its poleward edge (see Section~\ref{appendixF} and \fig{S5}). It is thus no surprise that we find no clear evidence of the jet on the northern edge of the oval. Within the framework of our simplified model, assuming a 300\,m/s counter-rotation wind inside the northern oval provides nonetheless a good fit to the measured wind speeds poleward of 55\degre N on the western limb where the northern oval was rising (see \fig{S3}). Finally, despite the northern aurora being located on the western side, mostly behind the terminator, we see a broad signal on the eastern limb at polar latitudes with an average LOS velocity of about +100\,m/s for which we lack a clear explanation. A more favorable observation geometry of the northern polar region is thus required to improve our understanding of the stratospheric circulation in this region.

\section{Discussion \label{Discussion}}
The branch of the northern auroral jet we tentatively detect is lying below the electrojet discovered at p $<$ 1\,$\mu$bar from infrared observations of H$_3^+$ emission by \citet{Rego1999} and further constrained by \citet{Stallard2001} and \citet{Johnson2017}. This electrojet has a near-to-supersonic velocity of $\sim$1-2 km/s and is in counter-rotation along the main oval \citep{Stallard2001,Stallard2003}. \citet{Achilleos2001} showed that the H$_3^+$ ions could accelerate the neutrals up to 60\% of their velocity through collisions between the ionosphere and the thermosphere in the ionization peak layer (0.07-0.3\,$\mu$bar). The upper limit set by \citet{Chaufray2011} of 1 km/s on the velocity of a corresponding H$_2$ flow confirmed a smaller neutral wind velocity, in agreement with our findings. Benefiting from ideal viewing conditions (sub-earth latitude of 0.2\degre N), \citet{Rego1999} also detected a similar counter-rotation electrojet on the main southern oval. Models by \citet{Majeed2016} and \citet{Yates2020} predict that neutrals have higher velocities below the southern oval than below the northern one. Although we find relatively similar velocities underneath the two ovals, our detection in the northern oval remains tentative such that we cannot conclude on the relative magnitude between the two auroral jets. This particular point thus needs to be confirmed with new observations. \citet{Majeed2016} and \citet{Yates2020} also predict that the southern jets are expected to disappear around the $\mu$bar level. On the contrary, our data demonstrate that the neutrals are still flowing with a substantial counter-rotation velocity at the sub-mbar level below the southern oval (and probably also below the northern one), i.e. $\sim$900\,km below the corresponding ionospheric winds of \citet{Rego1999} and 100-500\,km below the tentative H$_2$ flow of \citet{Chaufray2011}. Despite the strong signal-to-noise limitations of our CO observations at 3\,mbar, we find that the southern auroral jets are at least twice slower in the mbar range than at sub-mbar levels, possibly disappearing between the sub-mbar and the mbar levels.

The detection of these auroral vortices down to the sub-mbar level may bear crucial implications on Jovian atmospheric chemistry. The photolysis of CH$_4$ at the $\mu$bar level triggers the production of more complex hydrocarbons. The addition of energetic magnetospheric electrons, which are more abundant in the auroral region than anywhere else on the planet \citep{Gerard2014}, further favors this complex ion-neutral chemistry \citep{Wong2003}. The presence of auroral vortices down to the sub-mbar level could confine the photochemical products within this region, by preventing the mixing of the material inside the oval with the material outside, and thus increase the production of heavy hydrocarbons and aerosols. Auroral chemistry probably increases the production of C$_2$ species as already observed by \citet{Sinclair2018,Sinclair2019}, and the production of aerosols \citep{Zhang2013}. The counter-rotation direction of the wind in both ovals translates into a clockwise circulation on the northern oval and counterclockwise circulation on the southern one. Such circulation pattern, which appears to be similar to anticyclones in this respect, could induce subsidence interior to the auroral ovals \citep{Yates2020}. The photochemically produced species would then be transported downwards and could escape the auroral region at the mbar level where the vortices could be breaking up. This increased production of aerosols coupled to the downward motion could also result in the removal of HCN by adsorption onto the aerosol particles at pressures higher than 0.1\,mbar at auroral latitudes, as shown by our data. This adsorption mechanism was proposed for Titan by \citet{Anderson2016} and needs to be quantified under Jovian auroral conditions. Another effect of the downward motions would be adiabatic heating around the vortex break up level. Heating at the mbar level was observed inside both ovals by \citet{Sinclair2017} and could be the indication that this is actually the level at which the vortices break up. We note that the independence of this heating with respect to solar illumination conditions \citep{Sinclair2017} seems to disqualify aerosol heating as a cause. We see a sharp HCN emission increase in our data at the edges of the oval and it could indeed be proof of such heating between the oval and its surrounding region. However, the HCN line is not optically thick, and we cannot waive the degeneracy between a temperature and an abundance increase. 

The detection of stratospheric auroral jets in this work demonstrates that the Jovian atmospheric circulation is complex not only in the equatorial region owing to the QQO \citep{Cosentino2017,Giles2020,Antunano2020}, but also in its polar regions. Repeated observations with the northern aurora in the field-of-view are necessary for a better characterization of the counter-rotation stratospheric jet underneath the main oval similarly to the situation witnessed in the south.


\section*{Acknowledgements}
The authors thank P. Gratier for helping implement the MCMC method, and R. Johnson and T. Stallard for providing them with their infrared ionospheric auroral wind velocities. T.C. acknowledges funding from CNES and the Programme National de Plan\'etologie (PNP) of CNRS/INSU. This paper makes use of the following ALMA data: ADS/JAO.ALMA\#2016.1.01235.S. ALMA is a partnership of ESO (representing its member states), NSF (USA) and NINS (Japan), together with NRC (Canada), MOST and ASIAA (Taiwan), and KASI (Republic of Korea), in cooperation with the Republic of Chile. The Joint ALMA Observatory is operated by ESO, AUI/NRAO and NAOJ. 


\bibliographystyle{aa} 

\begin{appendix}   
\section{Observations, data reduction and imaging \label{appendixA}}
Observations of Jupiter for ALMA project 2016.1.01235.S were executed on March 22nd, 2017. We used 42 antennas of the 12-m telescope array. The shortest and longest baselines were 15.1 and 160.7\,m, respectively. The observations started with calibration observations between 4:46UT and 5:11UT. After initial pointing calibration, extended bandpass calibration observations were carried out on J1256-0547 to comply with the high spectral dynamic range required by the high S/N observations of CO and HCN emission lines on the bright Jovian continuum. Finally, amplitude calibration observations were acquired using Ganymede as a target. Between 5:11UT and 5:35UT, most of the observation time was spent on Jupiter, with regular phase calibration observations on J1312-0424. Jupiter had an average elevation of 72\degre~above the horizon and the sky transparency was adequate, with 0.85 to 0.95\,mm of precipitable water vapor. These conditions resulted in system temperatures at 345 and 354 GHz ranging from 120 to 190\,K and from 140 to 240\,K, respectively. We covered the full Jovian disk with a 39-point mosaic over a square region of 60\arcsec$\times$60\arcsec with Nyquist sampling.

The spectral setup enabled us to record simultaneously the Jovian emission of the HCN (5-4) line at 354.505\,GHz and CO (3-2) line at 345.796\,GHz. The HCN line was observed in two basebands with different bandwidths and spectral resolutions. The highest spectral resolution was 122.070\,kHz ($\sim$103\,m/s) over a bandwidth of 234.375\,MHz. The CO line was observed in a single baseband with a spectral resolution of 488.242\,kHz ($\sim$426\,m/s) over a bandwidth of 937.500\,MHz.

The data reduction process was run under CASA 4.7.2 and included pointing, bandpass, amplitude and phase calibrations. We also corrected for the relative radial velocity of Jupiter. We then produced continuum images in the different basebands. To produce the spectral data cubes in these basebands, we applied continuum subtraction before the imaging stage. Because of the limited uv-coverage in the short baselines, the extended emission of Jupiter is mostly filtered out and only the limb of the planet is imaged. We obtain a sensitivity of 48\,mJy/beam per 122\,kHz resolution element in the HCN baseband. With a peak line intensity of 1.2\,Jy/beam in the mid-to-low latitudes and peak line intensities up to 2.1\,Jy in the polar regions, we obtain detections with S/N ranging from 25 to 50 at 122\,kHz resolution depending on the latitude. In the CO baseband, we reach a sensitivity of 17\,mJy/beam per 488\,kHz resolution element and S/N ranging from 17 to 25.

\section{Contribution functions \label{appendixB}}
\begin{figure}[!h]
\begin{center}
   \includegraphics[width=8cm,keepaspectratio]{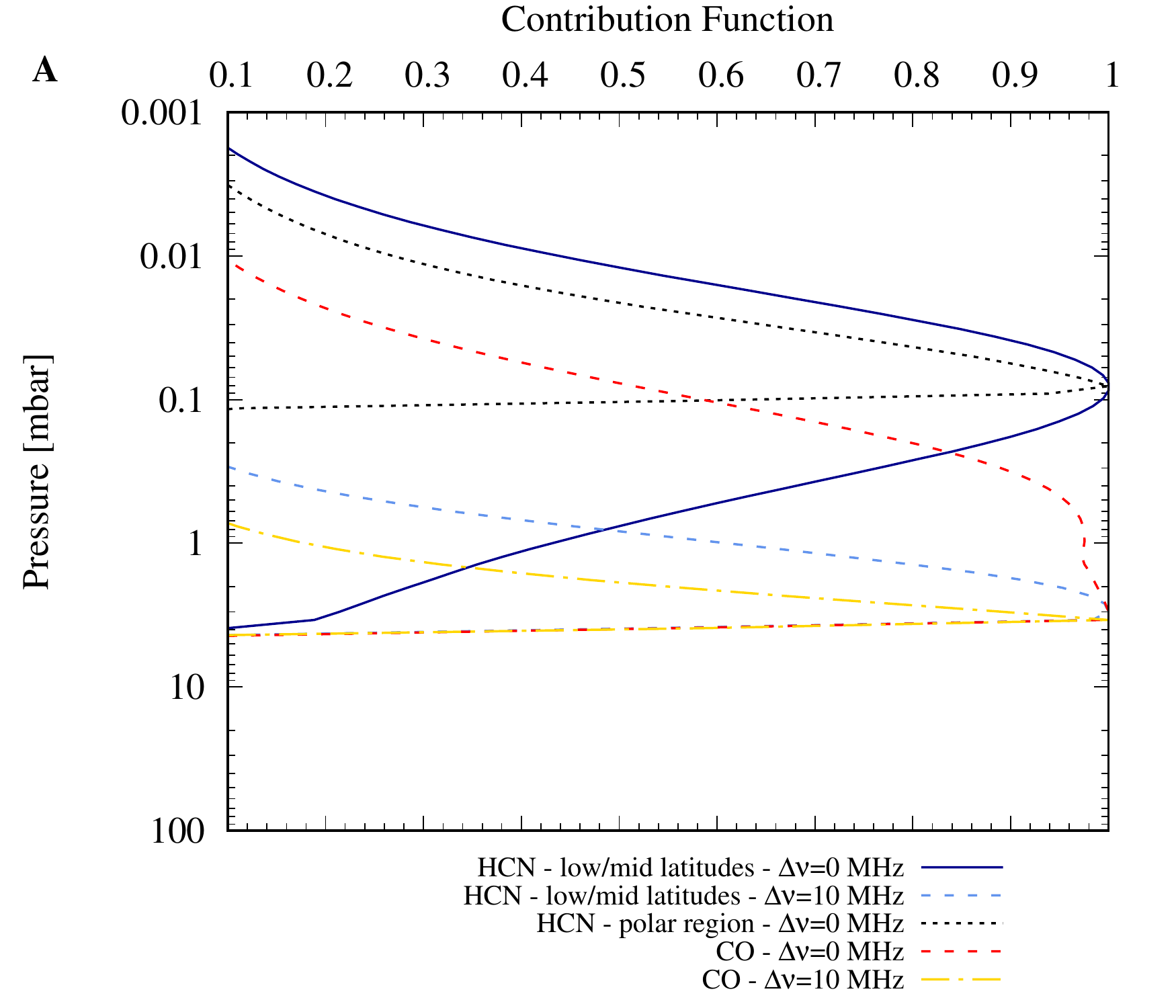}
   \includegraphics[width=8cm,keepaspectratio]{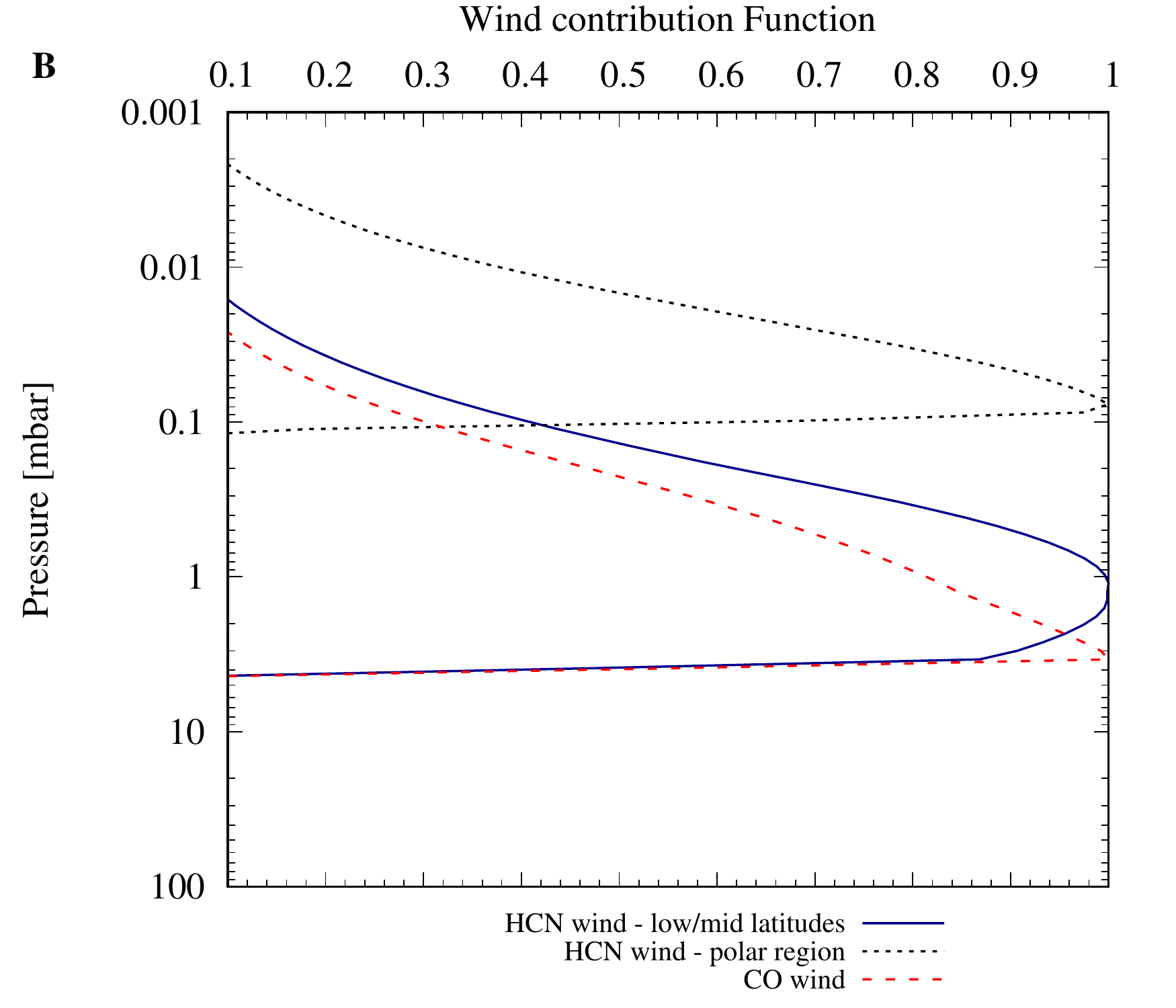}
\end{center}
\caption{Contribution and wind contribution functions. (A) Contribution functions of the HCN and CO lines at Jupiter's limb at the spectral and spatial resolutions of the observations. They are computed for different frequency offsets from the line center. For HCN, different locations (low/mid latitudes and polar regions) are presented. For CO, the contribution functions are similar whatever the latitude. (B) Wind contribution functions for HCN at low-to-mid (solid blue line) and polar latitudes (dotted black line), and for CO whatever the latitude (red dashed line). }
\label{S1} 
\end{figure}

We used step vertical profiles in which CO and HCN have a constant mole fraction $y_0$ above a cut-off pressure $p_0$. To reproduce the CO lines, which are essentially very similar in amplitude and width on the limb whatever the latitude, we set $p_0$ $= $5\,mbar and $y_0$ $=$ 4 $\times$ 10$^{-8}$. For HCN, we found $p_0$ $=$ 5 
\,mbar from 60\degre S to 50\degre N and $p_0$ $=$ 0.1\,mbar at polar latitudes. With $y_0$ $=$ 10$^{-9}$, we could fit most of the HCN lines, except within the auroral ovals where higher stratospheric temperature or a higher value for $y_0$ is required. From these radiative transfer simulations performed at the spatial and spectral resolutions of the observations, we derived the contribution functions of the CO and HCN lines at the limb, both at the line centers and in their wings (at 10\,MHz from the line center). The results shown in \fig{S1}-A indicate that the CO line center is formed at 0.5-5\,mbar levels, while its wings are formed around the 5-mbar level. We found that the HCN line center is formed at the $\sim$0.1\,mbar whatever the latitude. Outside polar latitudes, the line is much broader and its wings probe down to the 5-mbar cut-off level, like CO.

Following \citet{Lellouch2019}, we computed the ``wind contribution function'' of the HCN line to establish the levels probed by the winds (see \fig{S1}-B). We found that the HCN line enables measuring winds at $\sim$1\,mbar from 60\degre S to 50\degre N, and at $\sim$0.1\,mbar at polar latitudes. The CO line would, in principle, allow us to measure winds at 3\,mbar, but the S/N of the observations is insufficient given the spectral resolution and we can only set an upper limit on the wind speed at this pressure level. 

\section{Retrieval of wind speeds as a function of latitude \label{appendixC}}
Given the spatial resolution of our observations, spectral lines should be asymmetric in the case of a vertically varying wind profile. However, we note that the S/N is not sufficient to derive a full vertical wind profile. We therefore assume a vertically constant wind in the altitude layers where the lines are formed and study the meridional variability of the winds. We use a symmetrical parametrized line shape to fit the observations with an MCMC procedure. The profile is the following:
\begin{equation}
\begin{multlined}
f(A,\nu,\nu_0,\alpha,\beta,\gamma,\delta,\Gamma)=A\left(\frac{1}{\sigma\sqrt{2\pi}}\mathrm{e}^{{-\frac{1}{2}\left(\frac{\nu-\nu_0}{\sigma}\right)^2}}\right)^\delta \\
\times\left(\frac{2}{\pi\Gamma}\frac{1}{1+\left(\frac{\nu-\nu_0}{\Gamma/2}\right)^2}\right)^\gamma \left(\left|\nu-\nu_0\right|^\alpha + \frac{1}{(\beta\Gamma)^2 + (\nu-\nu_0)^2}\right) \\
\times\frac{1}{\Gamma^2 + (\nu-\nu_0)^2 + 1}
\end{multlined}
\end{equation}
The above expression is composed of four functions, and each one plays a role in shaping the line. The first one is a power Gaussian that enables reproducing the line core in the case of a strongly convolved line. The second one is a power Lorentzian meant for narrow line peaks.  The third one is the sum of a power absolute function and a modified Lorentzian that controls the amplitude of the line. The fourth is another modified Lorentzian to shape the line wings. $A$ is a constant.

The line profile is controlled by a set of eight parameters and $\nu$ is the frequency. Of these eight parameters, $\alpha$, $\beta$, $\sigma$ and $\delta$ are fixed to 6.0, 16.5, 0.202 (GHz) and 2.0, respectively, after numerical testing to obtain spectral lines with narrow peaks and narrow to broad wings like the HCN and CO lines of our dataset: $\alpha$ enables uplifting the line wings, $\delta$ enables producing narrow peaks and narrow wings, and $\beta$ mostly reduces the amplitude of the peak.

$A$, $\nu_0$, $\Gamma$ and $\gamma$ are MCMC fitting parameters, as well as the spectrum noise level. Some of them have pre-estimated ranges: $\Gamma\in\left[0.0001;0.1\right]$\,GHz, $\gamma\in\left[0.09;0.11\right]$, and $A$ is such that the amplitude of the line profile $f$ lies within 10\% of the amplitude of the observed line. Finally, $\nu_0$ is the central frequency of the line that we want to establish, and it combines the natural central frequency of the line, the rotation of the planet and the wind component projected on the line-of-sight. We estimate the fitting parameters (e.g. $A$ for the line amplitude) from each line before running the MCMC procedure.

We first performed a qualitative assessment of the fitting procedure and required computational time for each spectrum to find a good compromise between the number of ``walkers'', the number of iterations per ``walker'', and the total computational time. We found that convergence was reached after 540$\pm$100 iterations (also called burn-in size), such that we ensure convergence in an acceptable computational time in each case by setting a maximum of 2200 iterations for each of the 32 ``walkers''. The whole bandwidth is used in the fitting procedure. We selected the 30 best-fit obtained from these iterations to demonstrate in \fig{Map_spectra} the fitting quality obtained with our parametrized line shape.

\section{Systematic and random error analysis \label{appendixD}}
There are several sources of possible systematic and random errors at the various wind speed retrieval stages. The first, obvious, cause of uncertainty is the spectral noise. The fact that we use spectral resolutions of 103\,m/s for HCN may seem contradictory with our goal to detect winds with expected velocities of the order of 100\,m/s or less.  However, the observation of the full line profile with high S/N enables us to fit the whole line profile and derive the line center position with an accuracy that exceeds the spectral resolution by using tens to hundreds of spectral points in the fitting procedure. The uncertainty on the retrieved wind velocity $v_\mathrm{wind}$ can be estimated by the following empirical formula:
\begin{equation}
\Delta v_\mathrm{wind}\sim\frac{FWHM}{S/N\times\sqrt{FWHM/\Delta\nu}}\times\frac{c}{\nu_0}
\end{equation}
where $FWHM$ is the full-width at half maximum of the line, $S/N$ is the signal-to-noise ratio per independent spectral channel, $\Delta\nu$ is the spectral channel width, $c$ is the speed of light, and $\nu_0$ is the line rest frequency. For HCN, the lines have significantly different $FWHM$ between the low-to-mid and high latitudes: 8\,MHz and 2\,MHz, respectively. The resulting estimates of $\Delta\nu$ are $\sim$30\,m/s and $\sim$15\,m/s, respectively, and the MCMC fitting procedure gives $\Delta\nu$$\sim$20\,m/s and $\sim$10\,m/s, respectively.

Another source of random error is the continuum subtraction before the MCMC fitting procedure. Even though the HCN line is located in the far wing of the NH$_3$ line at 572\,GHz, the continuum should be flat within $\sim$0.1\% over the observed bandwidth. However, the continuum subtraction applied in the uv-plane within CASA before the imaging stage sometimes result in a non-flat continuum on the resulting spectra. We proceed with an additional subtraction of a 1st order polynomial from the spectra to flatten their continuum. This process implies a slight shift of the line center. This effect is relatively independent of the latitude and we estimate from our simulations that the added uncertainty on the velocity is lower than 10\,m/s on the HCN wind speeds. We add quadratically this uncertainty with the MCMC uncertainty.

\begin{figure}[!h]
\begin{center}
   \includegraphics[width=9cm,keepaspectratio]{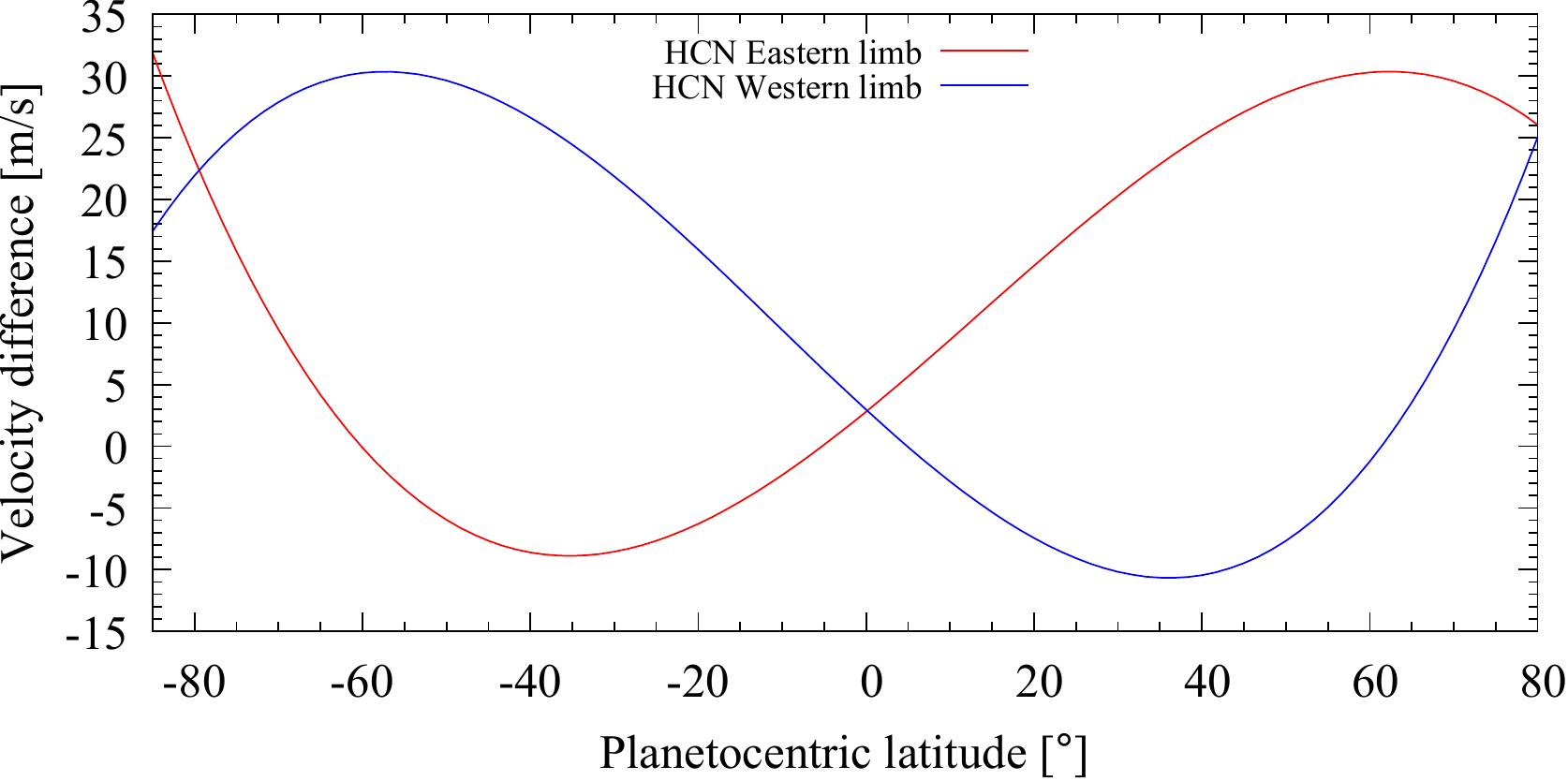}
\end{center}
\caption{Effect of pointing uncertainty on wind speed retrieval. Velocity difference as a function of latitude between two sets of LOS velocity retrievals: the one in which the planet is assumed to be centered in the field-of-view is subtracted from the other in which we applied the position offset as determined from the analysis of the continuum. Results are shown for both limbs. }
\label{S2} 
\end{figure}

At a given latitude, when the wind speed is lower than the noise level, one would expect the eastern and western limb wind speeds to be centered on the zero-velocity line. On the contrary, when a zonal wind is present, the east and west wind curves should in principle mirror each other. However, we initially found in the mid-latitudes, i.e. where there is no clear detection, that the curves suffered from a negative offset of $\sim$30\,m/s. In what follows, we list the sources of systematic errors that could be the cause of this offset, and detail how we treated them:
\begin{enumerate}[label=\alph*)]
  \item Jupiter radial velocity correction
The first obvious systematic error that could cause an overall shift of the wind speeds concerns the accuracy of the Jovian radial velocity correction. It is made within CASA by using JPL/Horizons ephemeris. The correction is applied at the level of individual integrations, i.e. every few seconds, by interpolating linearly between ephemeris table entries with the accuracy of the table. The geocentric radial velocity is tabulated with an accuracy of 10$^{-8}$\,UA/day, with one entry per day. By interpolating to a time within the range of our observations, we find that the systematic error on the radial velocity correction is $<$ 1\,m/s. 
  \item Offset of the planet center position in the spectral maps 
The wind-induced Doppler shift from each limb measurements are retrieved by subtracting the planetary rotation velocity projected along the line-of-sight to the fitted central frequency of the lines. It is important to account for the elliptical shape and orientation of the synthetic beam to compute the line shift induced by the planetary rotation.
We take the System III rotation period of Jupiter. To compute the spectral shift caused by the planet solid rotation for each extracted spectrum, we account for the projection of the rotation velocity vector on the LOS and the convolution by the synthetic beam.
After self-calibrating the data to center Jupiter at best on the image, there can still be a small center offset of up to 1/10 of a synthetic beam due to interferometric seeing. This affects our planet rotation speed subtraction process. To correct for this, we used the continuum images at 354\,GHz to constrain the location of the center of the planet. We found that the planet center was shifted by 53$\pm$5\,mas in right ascension and -32$\pm$6\,mas in declination. The curves presented in \fig{winds_summary} take these offsets into account. The remaining uncertainty equivalent to 5\,m/s is caused by continuum variability in the millimeter \citep{dePater2019}. This random error is added quadratically to the previous ones. When compared to the ideal case of a wind speed retrieval with a perfectly centered planet, this position shift results in a velocity difference as a function of latitude that is not uniform, because of the position angle of the planet. While an east-west equator-aligned offset shifts the two wind curves in the same direction, a north-south offset distorts the overall shifts. The difference is obtained by subtracting the two retrieved wind curves from one another, and it is shown in \fig{S2}. 
\end{enumerate}

\section{Modeling the spectral effect of a constant counter-rotation wind inside the auroral ovals \label{appendixE}}
To fit the northern and southern auroral wind speeds of \fig{winds_summary} (top), we simulated the Doppler-shifts induced by constant winds within the auroral ovals on the spectral lines. We took the radiative transfer model of \citet{Cavalie2019} and the HCN vertical profiles derived from the line shape analysis: from 60\degre S to 50\degre, HCN was set constant to 1\,ppb for $p_0$ $<$ 5\,mbar, and to zero at higher pressures. For latitudes lower than 60\degre S and higher than 50\degre N, HCN was set constant to 1\,ppb for $p_0$ $<$  0.1\,mbar, and to zero at higher pressures. For the auroral ovals, we took the oval inner and outer edges as defined by \citet{Bonfond2012} as a starting point and set constant wind speeds within these edges. We proceeded as follows:
\begin{enumerate}
  \item Simulation of spectral lines at infinite spectral and spatial resolutions on a highly sampled planetary grid. 
  \item Application of a spectral shift induced by the planet rotation for all LOS.
  \item Application of a spectral shift induced by the LOS component of the constant wind when the LOS intercepted the auroral oval.
  \item Convolution of the spectra spatially by the ALMA beam and spectrally for each studied pointing. 
  \item Measurement of the line center frequency for each studied pointing and subtraction of the contribution of the planet rotation, which was derived from a similar simulation in which there was no auroral wind.
\end{enumerate}

\begin{figure*}[!h]
\begin{center}
   \includegraphics[width=18cm,keepaspectratio]{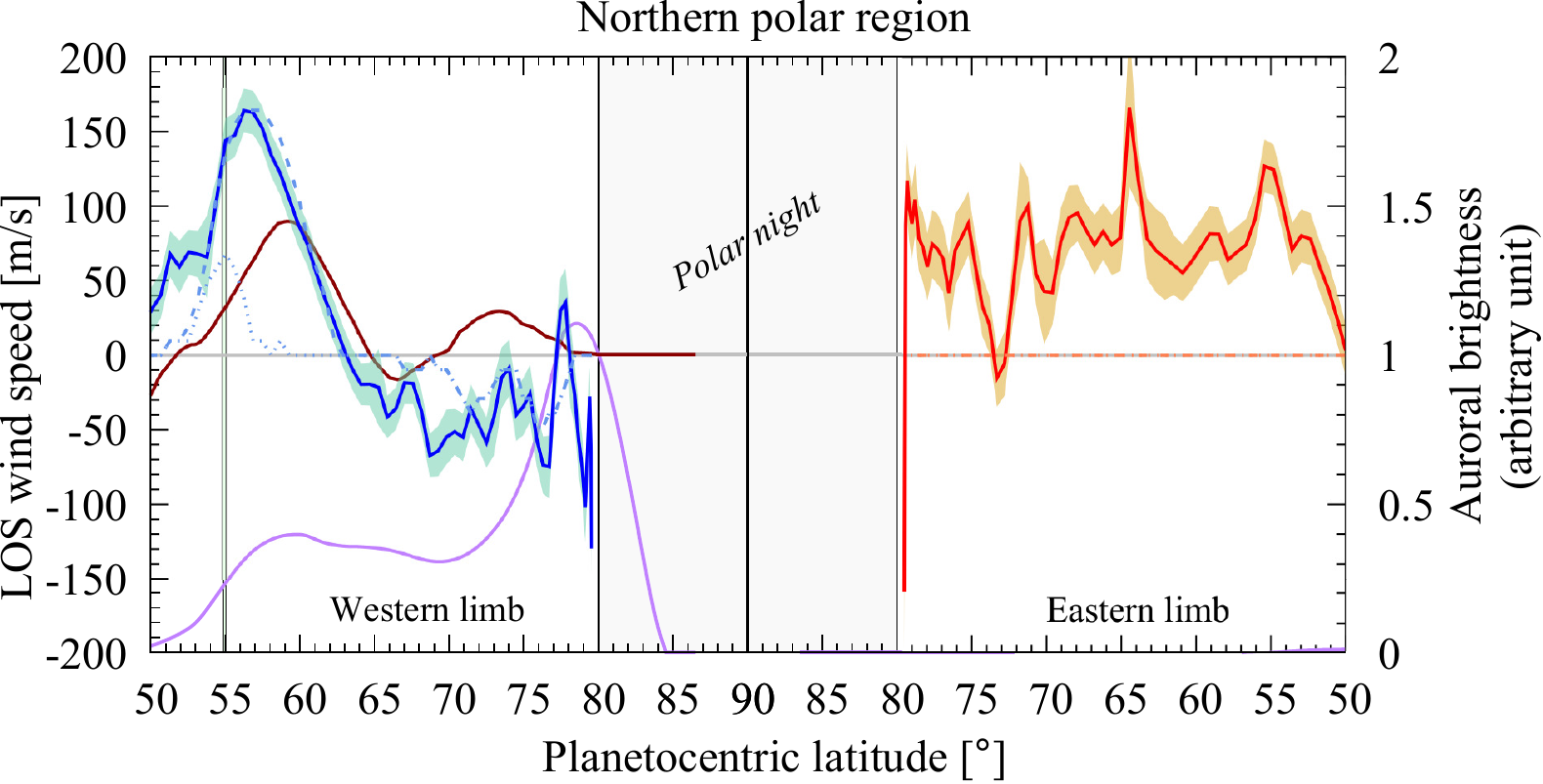}
   
   \vspace{0.5cm}
   
   \includegraphics[width=18cm,keepaspectratio]{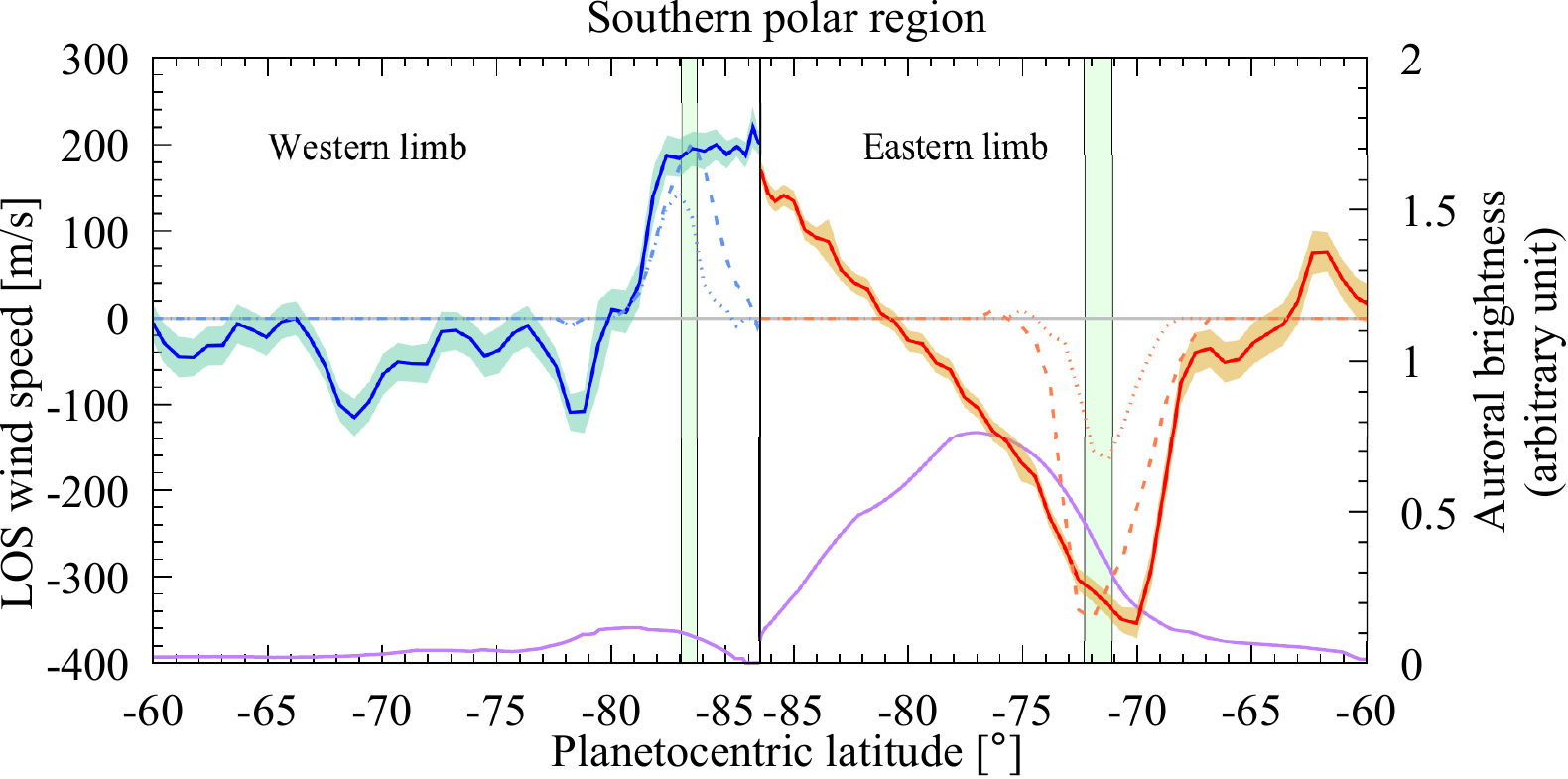}
\end{center}
\caption{Comparison of models with observations. Comparison between LOS velocities measured at 0.1\,mbar with ALMA (same color code as in \fig{winds_summary} top) and simulation results. A model with constant counter-rotation winds within the ovals as defined by \citet{Bonfond2012} results in too narrow wind peaks, as shown with the light blue and orange dotted lines, whatever the wind speed. The light blue and orange dashed lines are produced with constant counter-rotation winds within the auroral ovals of 300\,m/s (north) and 370\,m/s (south) in the case of extended ovals, as described in the main text. The latitudes swept by the M=30 footprints of the \citet{Connerney2018} magnetic field model is plotted with green stripes. The wind peaks are found within 1-2\degre~of these footprints. The statistical UV brightness model (purple line) derived from the observations of \citet{Clarke2009} and the ionospheric winds (brown line, speeds divided by 5 on the plot for an easier comparison) derived from the H$_3^+$ infrared observations of \citet{Johnson2017} are also included for a qualitative comparison with our measured wind speeds. Both UV brightness and H$_3^+$ wind curves have been degraded to the ALMA spatial resolution. The data from \citet{Johnson2017} are taken from their Figure 8-f. The northern auroral region is in the top panel, the southern one in the bottom panel.}
\label{S3} 
\end{figure*}

\begin{figure}[!h]
\begin{center}
   \includegraphics[width=9cm,keepaspectratio]{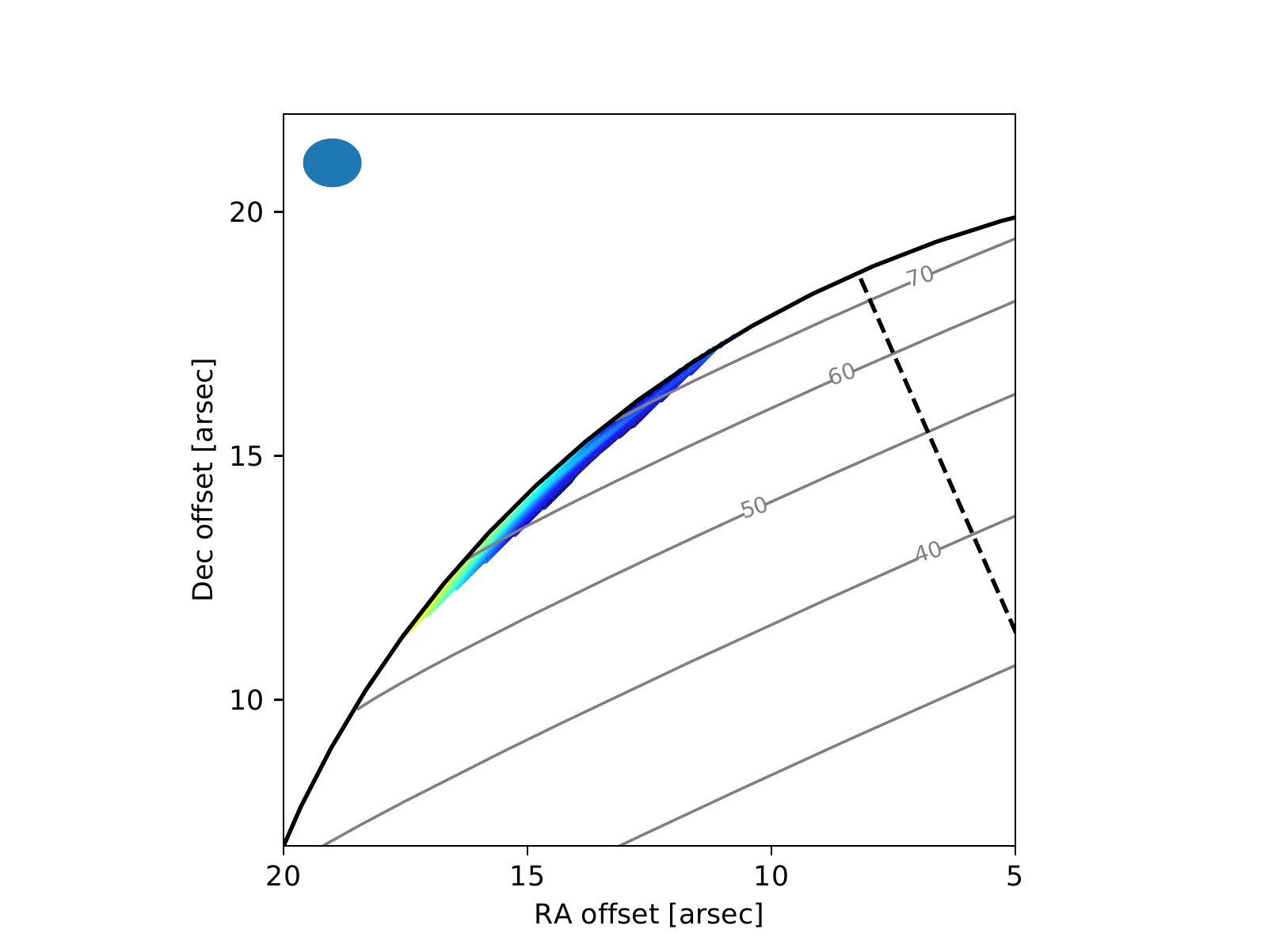}
   \includegraphics[width=9cm,keepaspectratio]{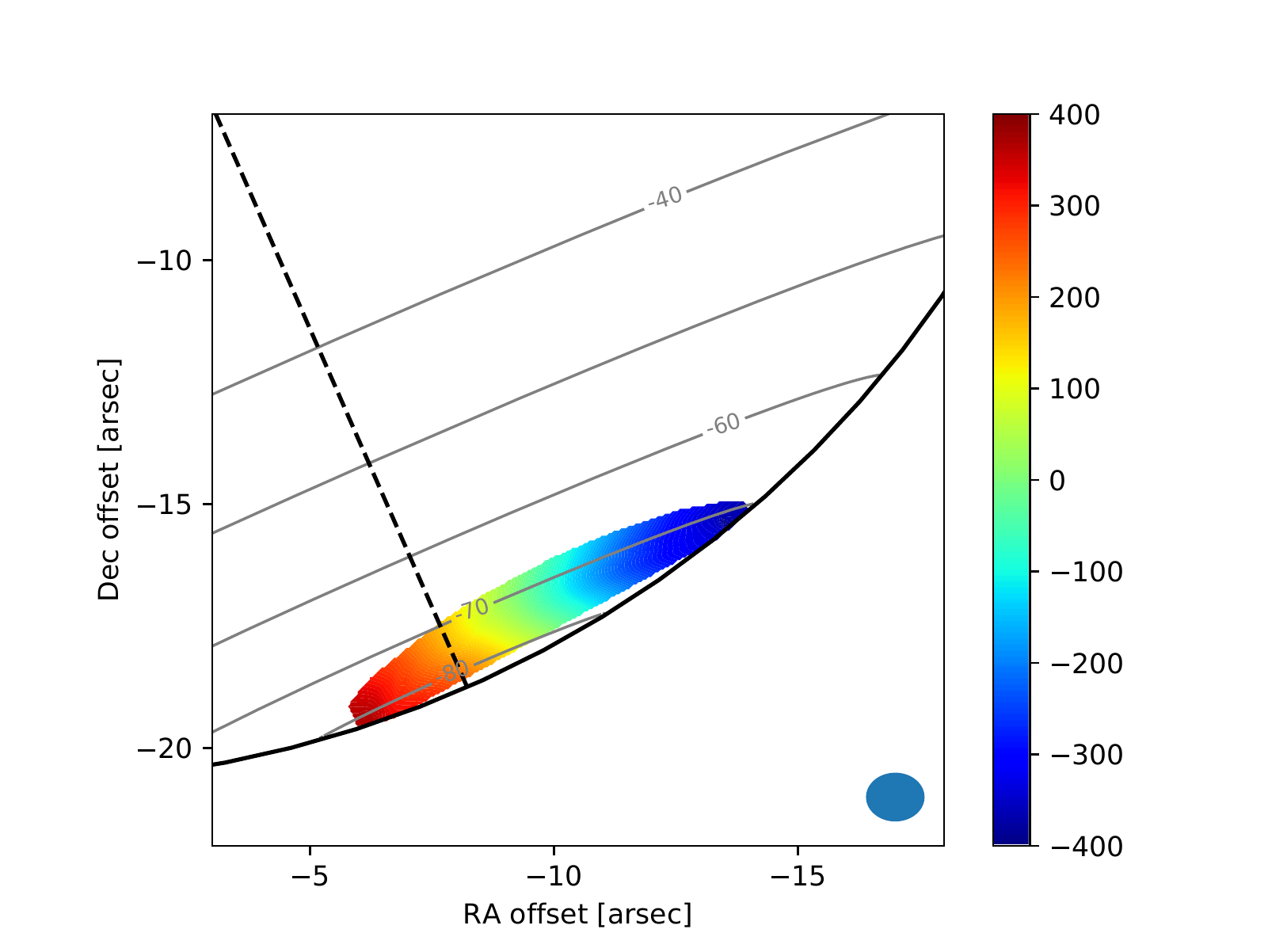}
\end{center}
\caption{Auroral oval wind model. LOS component of the winds in the auroral ovals at infinite spatial resolution, assuming a counter-rotation wind speed of 300\,m/s in the northern aurora (left) and 370\,m/s in the southern aurora (right). The ALMA synthetic beam size is displayed with a blue filled ellipse and the planetocentric latitudes are indicated.}
\label{S4} 
\end{figure}

On the south eastern limb, the oval is at -72\degre~and is $\sim$3\degre~wide in the model of \citet{Bonfond2012}. Using this definition of the oval does not provide a good fit to the data. The wind peak is too narrow as shown in \fig{S3}. The northern oval is also too narrow to provide the large wind peak seen at 57\degre N. We had to extend the southern oval by 1.5\degre~poleward and 2\degre~equatorward to improve the fit to the data. On the northern oval, and for similar reasons, we had to shift northward the southern inner edge by $\sim$5\degre. These adjustments of the size of the northern and southern ovals may translate from a more extended oval episode of the aurora. We found that wind speeds of 300\,m/s in the northern oval and 370\,m/s in the southern oval, both in counter-rotation, provide us with relatively good fit to the data given the simplicity of the auroral oval wind model. The LOS projected components of the auroral oval wind at infinite spatial resolution are presented in \fig{S4} and the comparison with the wind measurements is shown in \fig{S3}.

\section{Comparison with UV brightness distribution and IR ionospheric wind data at auroral latitudes \label{appendixF}}
\begin{figure}[!h]
\begin{center}
   \includegraphics[width=8cm,keepaspectratio]{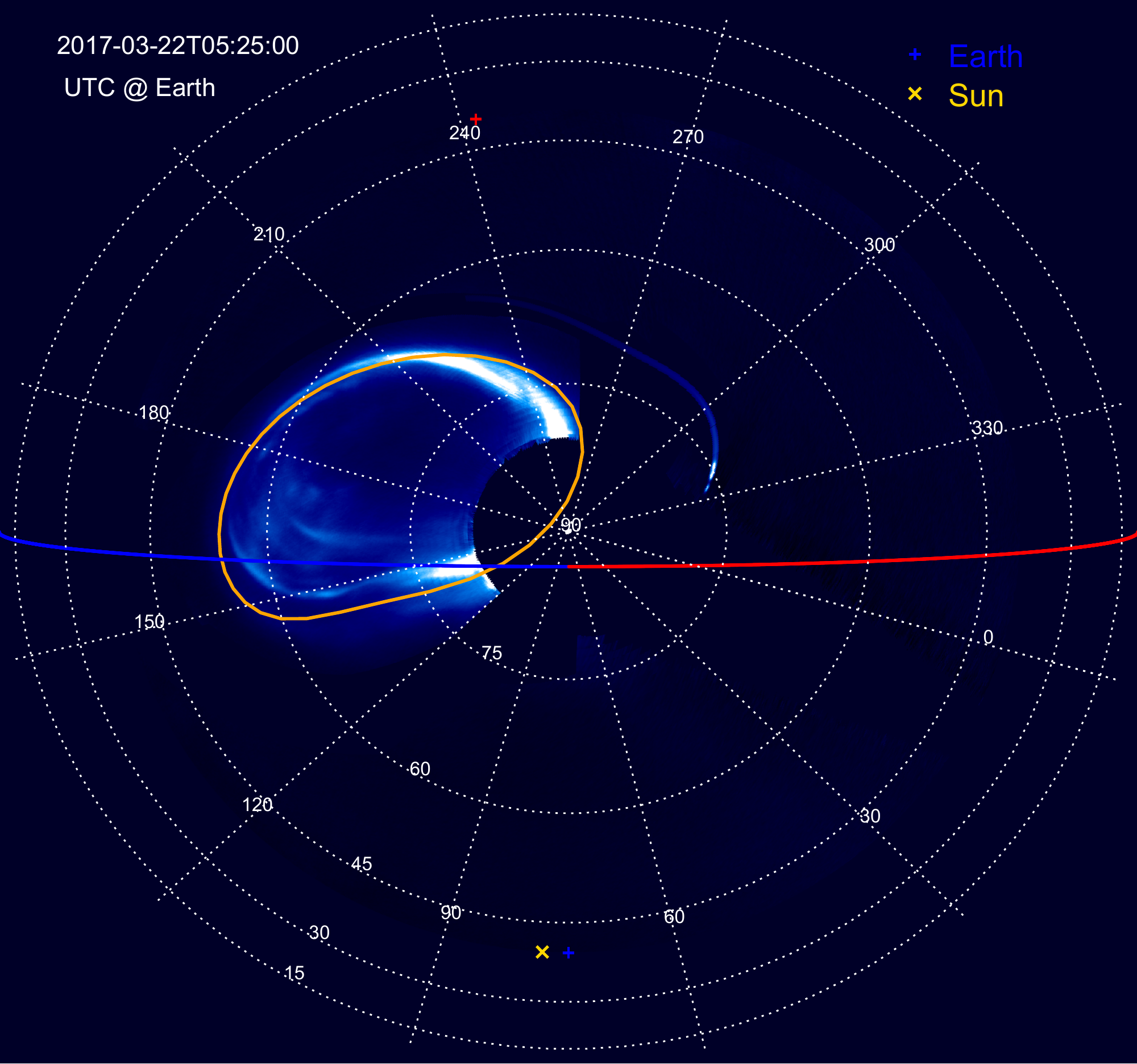}
   \includegraphics[width=8cm,keepaspectratio]{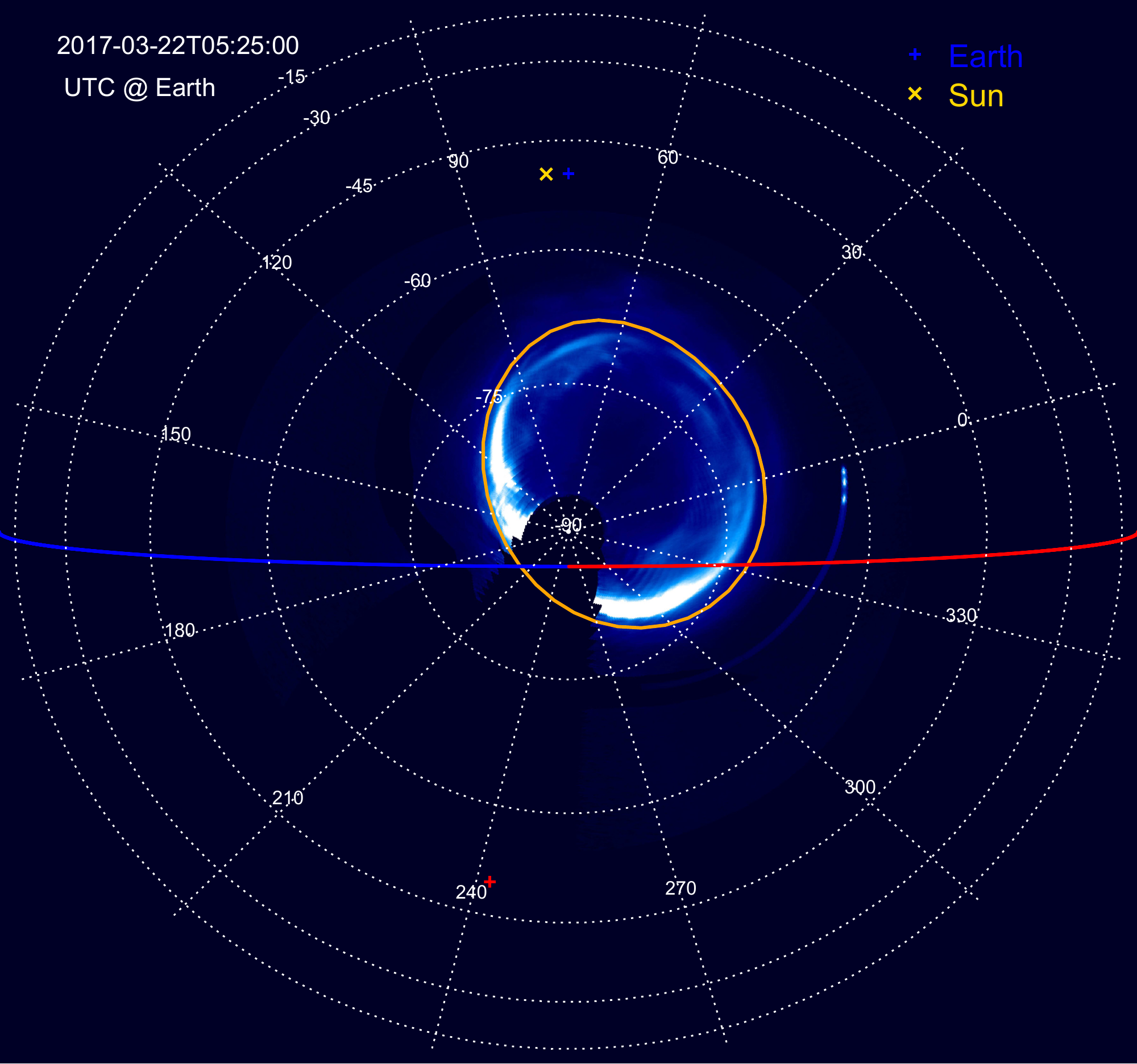}
\end{center}
\caption{UV aurora models at the time of the ALMA observations.   (Left) Northern polar projection of the brightness of the statistical UV aurora in the configuration of the ALMA observations. The red and blue lines represent the east and west terminators. The orange lines show the M=30 footprints of the magnetic field model of \citet{Connerney2018}. The sub-solar and sub-earth points are indicated with yellow and blue crosses, respectively. The red cross indicates the CML of the \citet{Johnson2017} measurements of ionospheric winds. (Right) Southern polar projection of the brightness of the statistical UV aurora in the configuration of the ALMA observations.}
\label{S5} 
\end{figure}

\fig{S3} also presents a comparison of the wind speeds measured at 0.1\,mbar in the polar regions with the brightness of the aurora seen in the UV. There were no UV observations performed simultaneously to our ALMA observations. We thus chose to use the statistical UV aurora brightness model derived from the observations of \citet{Clarke2009} and set it to the geometrical configuration of our ALMA observations, as presented in \fig{S5}. On the north western limb, the HCN wind peak at 57\degre N is relatively well collocated with the first UV peak (59\degre N) which corresponds to the southernmost part of the oval in our field-of-view. We see no obvious wind peak coinciding with northernmost part of the oval in the field-of-view, which also corresponds to the most intense UV peak (at 88\degre N), because this latitude is at the limit of the polar night region and wind measurements are therefore much noisier. At the south eastern limb, there is a larger offset of 7\degre~between the position of the oval, as determined from the UV brightness peak (77\degre S), and the peak seen in the wind speeds (70\degre S). Finally, we also find an offset of 2-3\degre~between the oval and the wind peak on the south western limb. The collocation of the wind speed peaks and the UV auroral brightness peaks of this statistical model are thus quite limited. And there seems to be no correlation between wind speed and UV brightness within the oval. However, the Jovian UV aurora have shown a significant degree of variability over all kinds of timescales (e.g. \citealt{Grodent2003}). The fact that we had to increase the width of the southern oval width to improve the fit to our data (see \fig{S3}) seems indicative of an episode of extended ovals.

Juno-UVS observations \citep{Gladstone2017} have shown that the position of the main ovals is well marked by the M$=$30 footprints of the magnetic field model of \citet{Connerney2018}. \fig{S3} shows that we observe the HCN polar wind peaks within 1-2\degre~of these footprints.

In \fig{S3}, we also compared our data with the ionospheric winds derived from H$_3^+$ observations by \citet{Johnson2017}. For this comparison to be as accurate as possible, we only took the data from their Figure 8-f; the CML was $\sim$240\degre, which places the position of the terminator within 10\degre~of the ALMA observation geometry. Here, we found a small offset of 3\degre~between the counter-rotation HCN and H$_3^+$ wind speed peaks, as shown in \fig{S3}. 

Because of the aurora variability, we need simultaneous observations to fully assess how well the stratospheric auroral oval winds detected in this work are collocated with the auroral UV brightness peaks associated with the ovals, and with the higher altitude auroral winds seen with H$_3^+$.

\end{appendix}

\end{document}